\documentclass[global,twocolumn,hyperref]{svjour-arXiv}
\usepackage{graphics}
\usepackage{here}
\usepackage{textgreek}
\usepackage{color}%カラー
\usepackage{cite}%cite番号順になる

\begin{document}
\title{Ungerminated Rice Grains Observed by Femtosecond Pulse Laser Second-harmonic Generation Microscopy\\
}
%\subtitle{Do you have a subtitle?\\ If so, write it here}
\author{\and{Yue Zhao}\inst{1}\hbox{\href{http://orcid.org/0000-0002-8550-2020}{\includegraphics{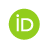}}} 
\and Shogo Takahashi\inst{1} \and Yanrong Li\inst{1} \and K. T. T. Hien\inst{1}\and Akira Matsubara\inst{1}\\　\and Goro Mizutani\inst{1}\thanks{\emph{Goro Mizutani:} mizutani@jaist.ac.jp}\hbox{\href{http://orcid.org/0000-0002-4534-9359}{\includegraphics{orcid.eps}}}
\and Yasunori Nakamura\inst{2}
}                     % Do not remove
%
%\offprints{}          % Insert a name or remove this line
%

\institute{School of Materials Science, Japan Advanced Institute of Science and Technology, Asahidai 1-1 Nomi, 923-1292, Japan \and Faculty of Bioresource Sciences, Akita Prefectural University, Akita City, Akita 010-0195, Japan}

\date{
\\
\\
{\it J. Phys. Chem. B}, {\bf 122(32),} 7855-7861 (2018). DOI:10.1021/acs.jpcb.8b04610}

\maketitle
\begin{abstract}
\bf As a demonstration that second-order nonlinear optical microscopy is a powerful tool for rice grain science, we observed second-harmonic generation (SHG) images of amylose- free glutinous rice and amylose-containing nonglutinous rice grains. The images obtained from SHG microscopy and photographs of the iodine-stained starch granules indicate that the distribution of starch types in the embryo-facing endosperm region (EFR) depends on the type of rice and suggests that glucose, maltose,
or both are localized on the testa side of the embryo. In the testa side of the embryo, crystallized glucose or maltose are judged to be detected by SHG. These monosaccharides and disaccharides play
an important role, as they trigger energy in the initial stage of germination. These results confirm SHG microscopy is a good method to monitor the distribution of such sugars and amylopectin in the embryo and its neighboring regions of rice grains.

\vspace*{-5mm}
\begin{figure}[h]
\centering
\begin{minipage}[t]{8.5cm}
\resizebox{1\textwidth}{!}{%
  \includegraphics{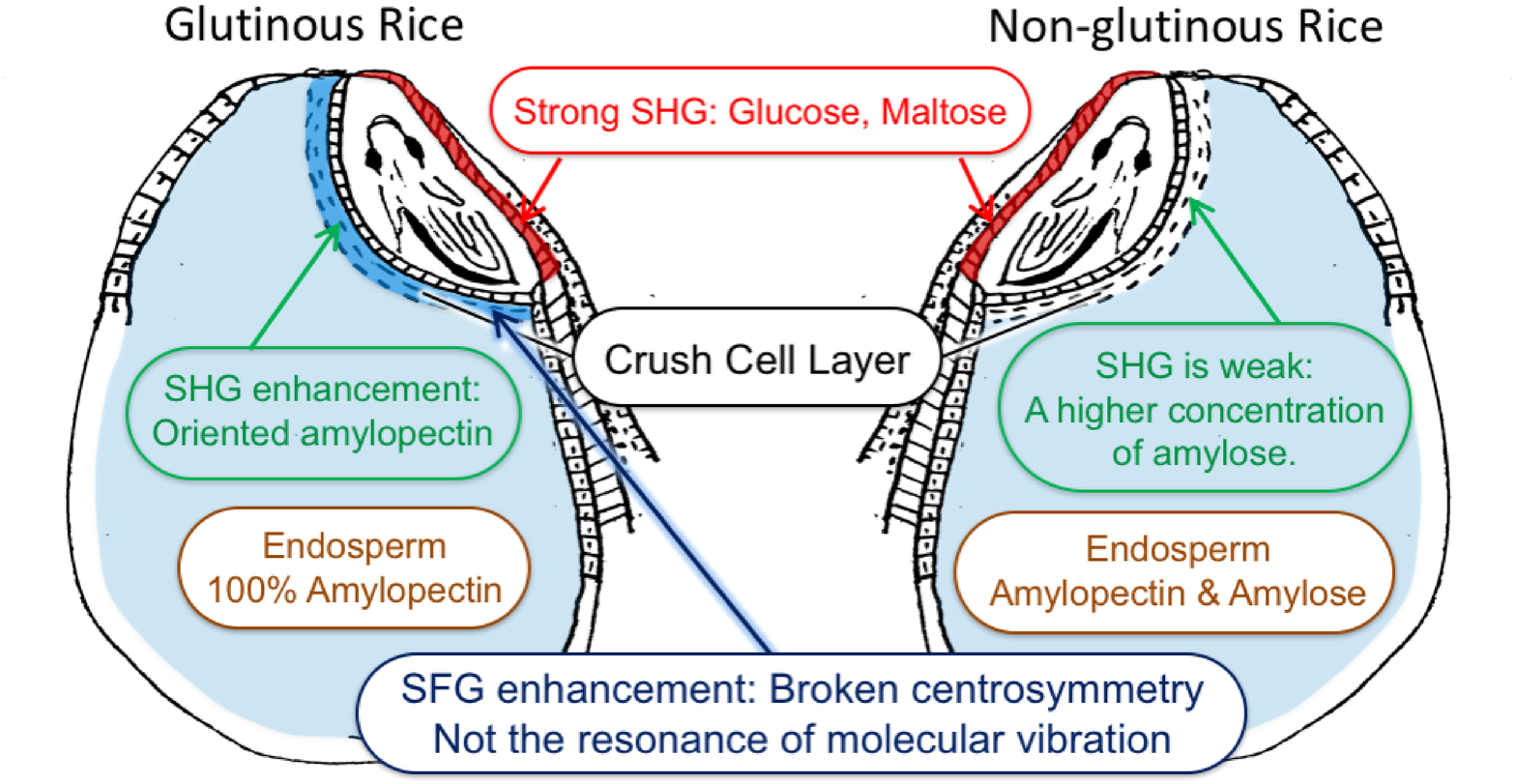}}
\end{minipage}%  
\vspace*{-15mm}
%\caption{Schematic illustration of the results of SHG in two varieties of rice. (a) The results obtained by SHG analysis in the cross section of the ungerminated glutinous rice.  (b) The results obtained by SHG analysis in the cross section of the ungerminated non-glutinous rice.}
\label{TOP-Fig6}      
\end{figure}
\end{abstract}
\vspace*{-5mm}
%

%\vspace*{5mm} 
\section{Introduction}
\label{sec:Introduction}

Rice is an important food. The endosperm of rice is mostly composed of starch, which is an energy source for most organisms. A previous study using sum frequency generation (SFG) microscopy reported the starch distribution in the cross section of a glutinous rice grain. The SFG signal is enhanced in the crush cell layer, which is an $\sim$100 \textmu m wide zone between the endosperm and the embryo \cite{li2012_3}.

Cells in the crush cell layer and the size of the starch granules are smaller than those in the center of the endosperm \cite{Inagaku_295}.  In the nuclear stage of endosperm formation, the endosperm nucleus grows remarkably in the vicinity of the embryo \cite{hoshigawa_1967a,hoshigawa_1972}. The embryo and the endosperm must interact and communicate at this stage because angiosperm seeds of overlapping fertilization ensure cooperative growth \cite{Lafon-Placette_2014}.  The embryo-facing endosperm region (EFR) develops differently from other endosperm tissues via programmed cell death \cite{hibara_2015}. Endosperm cells closest to the embryo do not accumulate starch granules. The next-nearest neighbor endosperm cells have special slender shapes, and the stored starch granules are small sized \cite{hibara_2015}.  Previous research has suggested that the construction of EFR and the size ratio of the embryo and endosperm are closely related \cite{hibara_2015}.  Therefore, morphological analysis in EFR is important for breeding science. On the other hand, during the stage of water absorption of germination, the crush cell layer acts as a water reservoir and absorbs water first \cite{takahashi1962_14}.  Then, starch is digested from the EFR near the scutellum as the endosperm develops \cite{Inagaku_75}. 

However, the origin of the SFG enhancement in the crush cell layer \cite{li2012_3} remains unclear. Two origins have been proposed for the enhanced large second-order nonlinear effect. One is the high density of amylopectin near the crush cell layer. The other is a high degree of orientation of amylopectin near the crush cell layer. To identify the origin, herein we observe the crush cell layers of two kinds of rice grains using SHG microscopy and map the broken symmetry of the molecular structure.

Optical second harmonic generation (SHG) occurs in non-centrosymmetric media within an electric dipole approximation, and its frequency is double that of the incident one. Hence, optical SHG microscopy can map asymmetrically oriented structures. The monosaccharide \textalpha -D-glucopyranose is a constituent unit of saccharide chains in starch. Since its structure is non-centrosymmetric, it displays a nonzero second- order nonlinear susceptibility and SHG activity. The anisotropic hydroxide and hydrogen bonds of aligned water have been reported as the origin of SHG in glucose and amylopectin hydrates \cite{OL_com_01}.  Additionally, D-glucose, D-fructose, and sucrose have been reported to show sum frequency generation (SFG) signals \cite{t1}.  

Starch, which is comprised of various components such as amylopectin and amylose, exhibits SHG activity \cite{32,cox2005_10}.  The origin of SHG in dry starch is amylopectin \cite{OL_com_02_12,zhuo2010_171,OL_com_03}.  Amylopectin has a tandem cluster structure where the sugar chains on small branches have a double helix structure \cite{s1-03_Kainuma1972}.  It is hardly soluble in water. Since its higher-order cluster structure has a macro- scopic asymmetry, the large nonlinear polarization in monosaccharide \textalpha -D-glucopyranose generates strong SHG light. In contrast, the amylose structure is speculated to be mostly amorphous in the starch granule of rice with a single helical structure \cite{takeda1986_148,hizukuri1989_189,manners1989_11,imberty1991_43,waigh1997_30}. The nonlinear polarization of the monosaccharide \textalpha -D-glucopyranose is thought to be canceled by the glucopyranose unit in the single helical structure. Thus, dry amorphous amylose does not display macroscopic nonlinear polarization, and has a very weak SHG \cite{OL_com_03,33}.  On the other hand, amylose can be crystallized in both single (V) and double (A and B) crystal forms \cite{OL_com_05,OL_com_06}. The ordered hydrogen and hydroxyl bond networks in the hydrates of these crystals are reported to allow SHG \cite{OL_com_04}. 

Generally, a starch iodine reaction is an easy and useful method to observe the starch distribution. However, the sample may permeate, resulting in an irreversible change. Although the method of F. Vilaplana {\it et al.} can quantify amylopectin and amylose in starch \cite{OL_com_07}, it cannot map their densities.

Second-order nonlinear optical microscopy is a non-destructive method. Amylopectin in starch generates an SHG signal due to macroscopic asymmetry \cite{OL_com_02_12,32,zhuo2010_171}, but amorphous amylose does not show SHG \cite{OL_com_03,33}. Thus, the distribution of amylopectin in the endosperm of amylose-free glutinous-type rice cultivar can be examined using SHG microscopy. In the initial stage of germination of a rice seed, the morphological structure of the crush cell layer, which is a special structure between the embryo and the endosperm that serves as a water reservoir, and the distribution of monosaccharides and disaccharides, which trigger energy in the embryo, have yet to be accurately examined.

Herein, SHG images of a typical nonglutinous rice cultivar Koshihikari, which includes both amylose and amylopectin, and a typical glutinous rice cultivar Shintaishomochi, which includes only amylopectin, are obtained using SHG microscopy with a femtosecond pulse laser and a morphological analysis. The comparison of the two cultivars confirms that amylose does not affect the experimental results. To our knowledge, previous research has not directly compared these two cultivars, demonstrating that SHG microscopy is a powerful tool to evaluate the distribution of sugars and amylopectin in the embryo and its neighboring regions of rice grains.

%%%%%%%%%%%%%%%%%%%%%%%%%%%%%%%%%%%
%%%%%%%%%%%%%%%%%%%%%%%%%%%%%%%%%%%
%%%%%%%%%%%%%%%%%%%%%%%%%%%%%%%%%%%

\section{Experimental Setup and Method}
\label{sec:Experimental Setup and Method}

\begin{figure}[hb]
%\vspace*{-6mm}
\centering
\begin{minipage}[t]{8.5cm}
\resizebox{1\textwidth}{!}{  \includegraphics{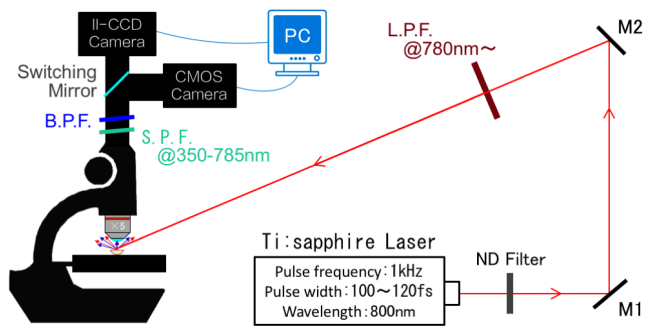}}
\end{minipage}%
\vspace*{1mm}  
\caption{Experimental setup for SHG microscopic observation using an II-CCD camera, femtosecond Ti:sapphire pulse laser, and regenerative amplifier. Details are described in the main text. M1, M2: Mirror. L.P.F.: Long wavelength pass filter. S.P.F.: short wavelength filter. B.P.F.: bandpass filter.}
\label{1}  
\end{figure}

In the experimental setup (see Fig. \ref{1}), the light source was a titanium sapphire femtosecond pulsed laser (Spectra-Physics: Tsunami) with a regenerative amplifier (Spectra-Physics: Spitfire) with a repetition frequency of 1 kHz, pulse width of 120 fs, and wavelength of 800 nm. The beam directly irradiated the sample without a focusing lens. The irradiated area on the sample stage with the sample was about 170 mm$^2$. The incident light energy density per unit area on the sample was 0.4 nJ/\textmu m$^2$, and the incident angle with respect to the normal to the sample stage was 60$^\circ$. A CMOS camera (Lumenera corporation, Lu135M) was used to monitor the sample surface using an incandescent lamp. The images of the scattered SHG light were observed by microscope optics (OLYMPUS BX60) with an image intensified charge coupled device (II-CCD) camera (HAMAMATSU, PMA-100). The power of the excitation light on the sample surface was controlled by using ND (Neutral Density) filters.

The sample was placed on a silicon wafer substrate. The SHG intensity of the oxidized Si surface is very weak compared to the SHG of bulk amylopectin rice seeds, and thus, it is negligible. Before the sample, a long wavelength pass filter (L.P.F.) was placed to block light with wavelengths shorter than 780 nm while allowing the 800 nm wavelength beam to pass. The scattered SHG light from the sample passed through an objective lens with a $\times$5 and NA of 0.15, became a parallel ray, and then passed through a short wavelength pass filter (S.P.F.) with a transmission wavelength range of 350−785 nm. This short pass filter blocked the 800 nm wavelength light. Finally, the SHG signal was selected by Semrock’s (FF01-395/ 11) band-pass filter (B.P.F.) with a center wavelength of 395 nm (the transmittance at 400 nm wavelength is over 90\%), and two-photon excitation fluorescence (2PEF) was selected by Semrock’s (FF02-438/24) band-pass filter with a center wavelength of 438 nm. Both the S.P.F. and B.P.F. were tilted 5$^\circ$ with respect to the wavefront of the parallel ray to eliminate \textquotedblleft ghost\textquotedblright signals. The spatial resolution of the microscope was determined by the chip size of the II-CCD camera (11 \textmu m $\times$ 13 \textmu m). In this case, the resolution was 2.6 \textmu m for a 5$\times$ magnification. The imaging integration time was 60 s. The polarization of the incident light was parallel to the horizontal direction of the microscopic images of Figs. \ref{2}(e), \ref{3}(e), \ref{4}(c) and \ref{5}(c).  The polarization of the detected SHG was not chosen.

The SHG signal was identified from the difference in the intensity distributions between 400 and 438 nm. There are two possible origins for the 400 nm signal in an image with a 400 nm wavelength bandpass filter. One is an SHG signal. The other is multiphoton excited fluorescence. In the latter case, the sample is excited by a multiphoton transition and emits luminescence pulses at a photon energy lower than the multiphoton energy. If the signal is dominated by the latter, observation wavelengths of 400 and 438 nm should provide similar results, and the images for both B.P.F.’s should be similar, because the spectra of multiphoton-excited fluorescence should be continuous near the two-photon wavelength \cite{2PEF_plant_3}.  If the spatial distributions of the signals clearly differ, then the signal from the 400 nm B.P.F. is SHG, whereas that for the 438 nm B.P.F. is 2PEF. 

In this study, the maximum incident light energy density per unit area on the sample was 0.4 nJ/\textmu m$^2$. No damage was observed at this incident light energy density. When observing starch by a scanning type SHG microscopy \cite{OL_com_01,OL_com_04}, the excitation beam was focused on the sample via an objective lens. In this case, the irradiation area should be 0.23 \textmu m$^2$ for a wavelength of 800 nm and a numerical aperture (NA) of the focusing lens of 0.75. When the energy of one pulse was 2 nJ, the incident light energy density per unit area was 8.7 nJ/\textmu m$^2$ \cite{OL_com_01,OL_com_04}.  This comparison shows that the current method is much less likely to damage the sample than typical scanning SHG microscopy. The nonscanning type setup can give a wider field of view than the scanning type SHG microscope, so it is convenient for samples with a large size such as the rice seeds. Our SHG microscope can acquire images in a short time such as several seconds to several minutes. This is because the excitation light energy density at the sample surface is made comparable to that of the scanning type microscope by using the regenerative amplifier. The scanning type laser microscope can give a diffraction limit spatial resolution, and it is better than the typical resolution of our type.
%%%%%%%%%%%%%%%%%%%%%%%%%%%%%%%%%%%

%\vspace*{1cm} 
\section{Results}
\label{sec:Results}

\begin{figure*}[t]
\centering
\begin{minipage}[t]{16.6cm}
\resizebox{1\textwidth}{!}{%
  \includegraphics{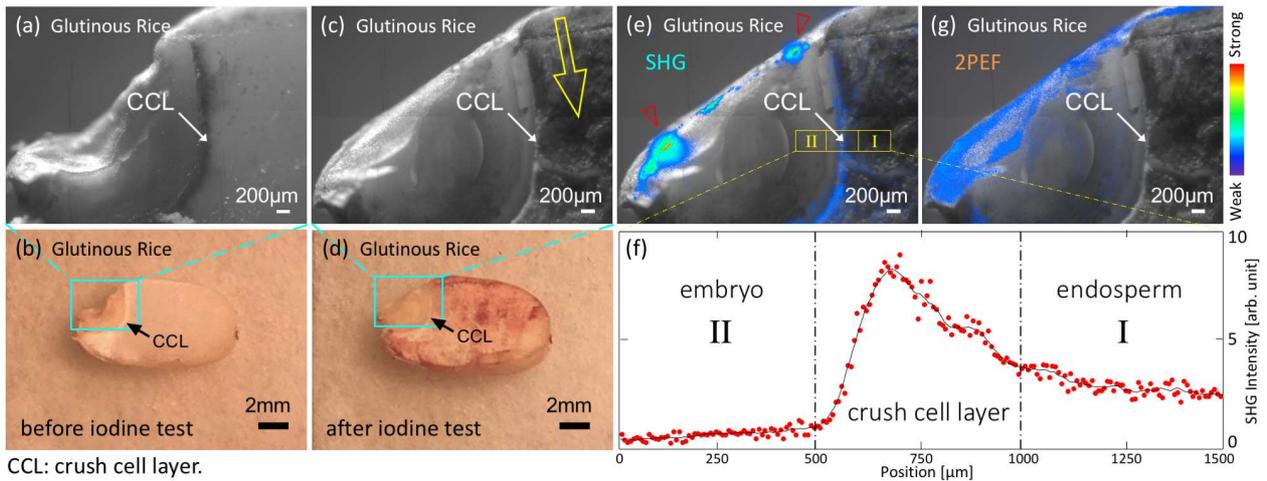}}
\end{minipage}% 
%\vspace*{1mm} 
\caption{Ungerminated glutinous rice ({\it {Oryzaglutinosa}} cv. Shintaishomochi) grain images. (a) Microscopic image of the glutinous brown rice cross section and (b) a photograph of the whole grain cross section. (c) Microscopic image and (d) macroscopic photograph after starch iodine reaction for 1 h. (e) SHG image of the same grain as in part c. (f) SHG intensity distribution within the yellow frame in part e. (g) 2PEF image of the same grain as in part c. The cyan frames in photographs b and d show the field of view of the microscope images of parts a, c, e, and g. Linear images a and c are illuminated by white light. SHG (e) and 2PEF (g) images are overlaid on the corresponding linear microscopic images. It is pseudocolored by the intensity scale bar at the right. In part f, each intensity is the summation of the corresponding vertical pixel SHG intensities. Note the strong SHG spots in the embryo near the embryo testa and the enhanced SHG at the crush cell layer.}
\label{2}      
\end{figure*}

\begin{figure*}[h]
\centering
\begin{minipage}[t]{16.6cm}
\resizebox{1\textwidth}{!}{%
  \includegraphics{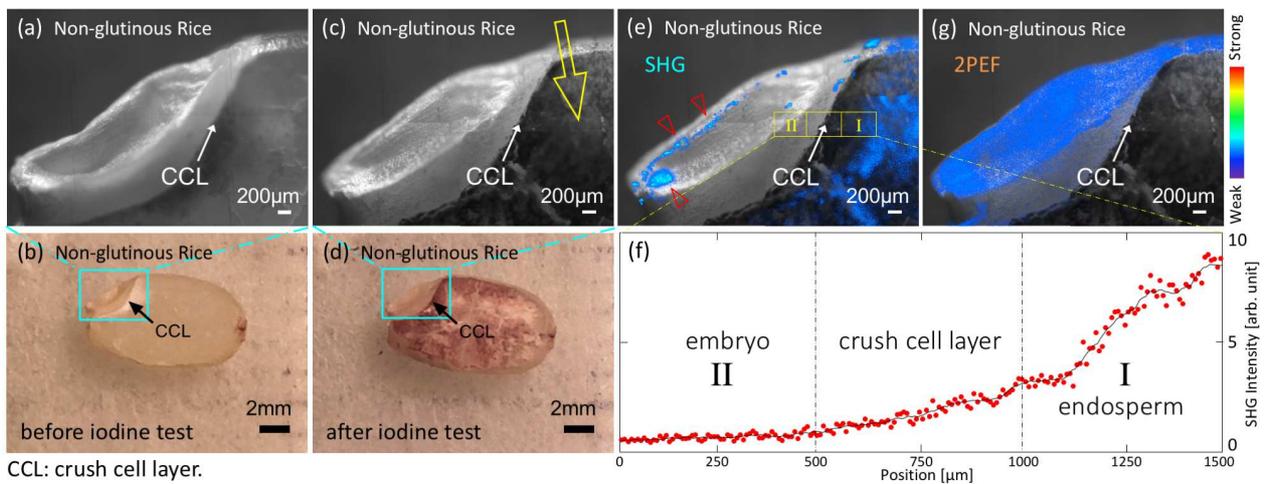}}
\end{minipage}%  
%\vspace*{1mm} 
\caption{Ungerminated nonglutinous rice ({\it {Oryzaglutinosa}} cv. Koshihikari) grain images. (a) Microscopic image of the nonglutinous brown rice cross section and (b) a photograph of the whole grain cross section. (c) Microscopic image and (d) macroscopic photograph of 5 min starch iodine reaction. (e) SHG image of the same grain as in part c. (f) SHG intensity distribution within the yellow frame in part e. (g) 2PEF image of the same grain as in part c. The cyan frames in photographs b and d show the field of view of the microscope images of parts a, c, e, and g. Linear images a and c are illuminated by white light. SHG (e) and 2PEF (g) images are overlaid on the corresponding linear microscopic images. It is pseudocolored by the intensity scale bar at the right. In part f, each intensity is the summation of the corresponding vertical pixel SHG intensities. Note the strong SHG spots in the embryo near the embryo testa and a gradual change of the SHG intensity at the crush cell layer.}
\label{3}      
\end{figure*}

First, the grain cross section of ungerminated brown rice was observed. Figures \ref{2} and \ref{3} show grains from amylose-free glutinous rice cv. Shintaishomochi and amylose-containing nonglutinous rice cv. Koshihikari, respectively. The photographs before and after the starch-iodine staining of the cross section of glutinous rice are shown in Fig. \ref{2}(b) and (d), while parts (b) and (d) of Fig. \ref{3} show those of nonglutinous rice, respectively. Figures \ref{2}(a), \ref{2}(c), \ref{3}(a) and \ref{3}(c) are their expanded images taken by the CMOS camera. The SHG images of the corresponding samples are shown in Figs. \ref{2}(e) and \ref{3}(e). The SHG image intensity is pseudocolored by the intensity scale bar on the right and superimposed on the images in Figs. \ref{2}(c) and \ref{3}(c). Because the iodine solution does not affect the SHG intensity under the current conditions \cite{32}, the SHG images after the addition of iodine solution are shown. 

Cisek {\it et al.} \cite{OL_com_04} reported that a hydration treatment alters the SHG in glucose, amylose, and starch. However, we found that the water absorption has a negligible impact on the relative SHG intensity in the images. Kong {\it et al.} \cite{Kong2014} reported that amylose in the presence of small guest molecules of iodine forms 6-fold left-handed single helices packed in an antiparallel arrangement and shows SHG activity. They oriented the amylose molecules by pulling the amylose film before and during exposure of the film to iodine gas \cite{Bluhm1981}. This study does not apply a mechanical force on the sample. Thus, the amylose remains amorphous in the rice grains, and the contribution of amylose to the SHG response is regarded as negligible. The signals in Figs. \ref{2}(e) and \ref{3}(e) are SHG because the images in Fig. \ref{3}(e) and (g) differ from each other.

The images in Figs. \ref{2}(c) and \ref{3}(c) show the two kinds of rice cross sections expanded by water absorption in the starch iodine test. These images slightly differ from those in Figs. \ref{2}(a) and \ref{3}(a). First, the iodine reagent, which reacts more strongly to amylose than amylopectin, colors the endosperm of both the glutinous rice and the nonglutinous rice (Figs. \ref{2}(d) and \ref{3}(d)). Figures \ref{2}(c) and \ref{3}(c) are black and white images, but the yellow hollow arrows denote the colored parts. In the glutinous rice grains ({\it {Oryzaglutinosa}} cv. Shintaishomochi) in Fig. \ref{2}(d), the color of the endosperm near the crush cell layer is weaker, suggesting that the crush cell layer either lacks or has a small amount of amylopectin. On the other hand, in the nonglutinous rice grains ({\it {Oryzaglutinosa}} cv. Koshihikari) in Fig. \ref{3}(d), the coloring of the endosperm near the crush cell layer is stronger than the other part, indicating that amylose is present at higher densities in that region.

%\vspace*{3mm} 

\begin{figure*}[h]
\centering
\begin{minipage}[t]{16.6cm}
\resizebox{1\textwidth}{!}{%
  \includegraphics{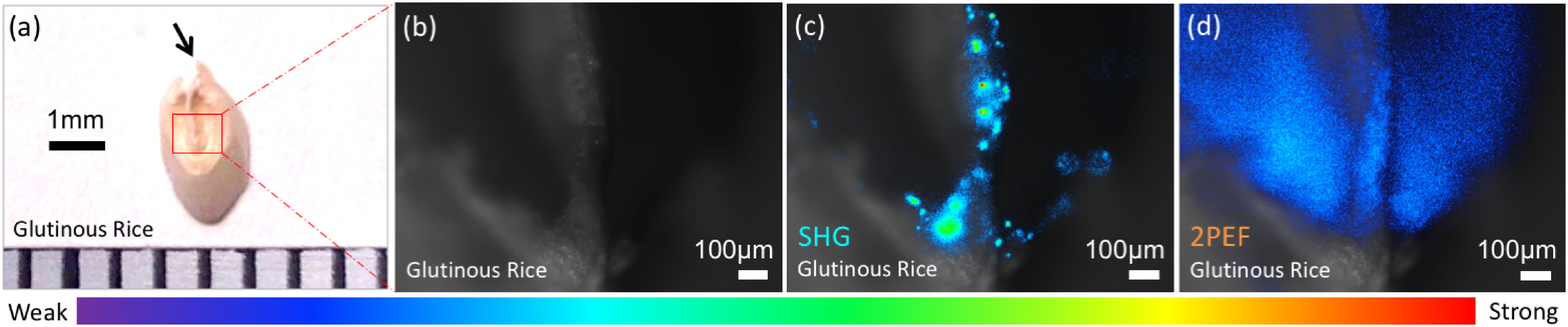}}
\end{minipage}%
%\vspace*{5cm}    
\caption{Images of ungerminated glutinous rice ({\it {Oryzaglutinosa}} cv. Shintaishomochi) grains. (a) Macroscopic photograph from the embryo side. (b) Linear microscopic image. (c) SHG image. (d) 2PEF image. Linear images are obtained by illuminating the sample by white light. SHG (c) and 2PEF (d) images are overlaid on the respective linear microscopic images. The SHG image is pseudocolored according to the intensity scale bar at the bottom.}
\label{4}      
\end{figure*}

%\vspace*{1cm} 

\begin{figure*}[h]
\centering
\begin{minipage}[t]{16.6cm}
\resizebox{1\textwidth}{!}{%
  \includegraphics{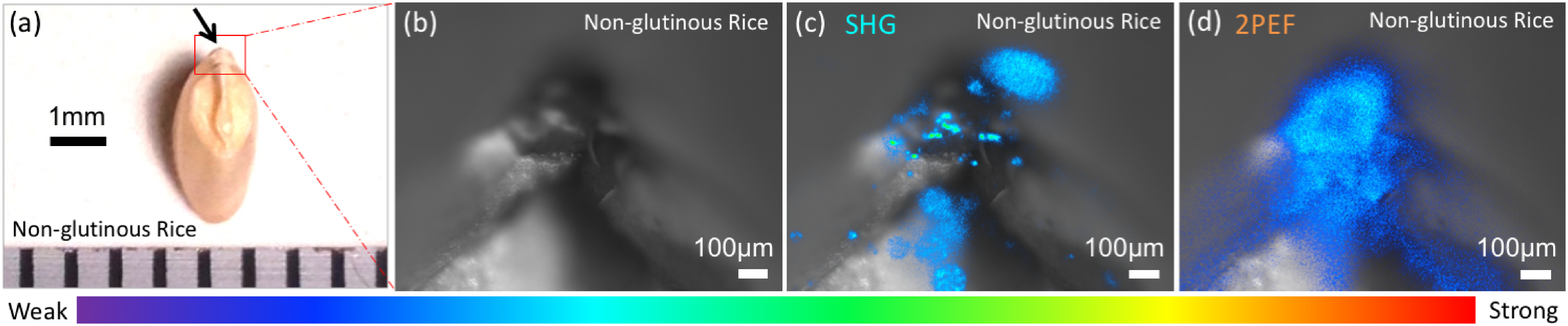}}
\end{minipage}%
%\vspace*{5cm}    
\caption{Images of ungerminated nonglutinous brown rice ({\it {Oryzaglutinosa}} cv. Koshihikari) grains. (a) Macroscopic photograph from the embryo side. (b) Linear microscopic image. (c) SHG image. (d) 2PEF image. Linear images are obtained by illuminating the sample by white light. SHG (c) and 2PEF (d) images are overlaid on the respective linear microscopic images. The SHG image is pseudocolored according to the intensity scale bar at the bottom.}
\label{5}      
\end{figure*}

%\vspace*{1cm} 

\begin{table*}[h]
\centering
\caption{Results of SHG Microscopy of the Dried Residue}
\label{t1}
\begin{tabular}{lllll}
\hline
sample & \multicolumn{1}{c}{not washed} & \multicolumn{1}{c}{water washed} & \multicolumn{1}{c}{DMF washed} & \multicolumn{1}{c}{ethanol washed}  \\
SHG    & active  & inactive & inactive   & active\\
\hline
\end{tabular}
\end{table*}

The SHG signals are observed from the endosperms of both the glutinous and the nonglutinous rice (Figs. \ref{2}(e) and \ref{3}(e)).  In the crush cell layer, the SHG signal is enhanced in the glutinous rice (Figs. \ref{2}(e, f)), but not in the nonglutinous rice (Figs. \ref{3}(e, f)).

\vspace*{-1mm}

SHG signals along the testa edge of the embryo exhibit a very strong intensity (Figs. \ref{2}(e) and \ref{3}(e), empty red arrow heads). To further investigate the strong SHG signals from the testa edge of the embryo in Figs. \ref{2}(e) and \ref{3}(e), the outer walls on the testa side of the embryo of the two brown rice grains were observed from the outer side (Figs. \ref{4} and \ref{5}).  Very strong SHG spots are seen along the hypocotyl in Figs. \ref{4}(c) and \ref{5}(c). A 2PEF signal is observed (Figs. \ref{4}(d) and \ref{5}(d)), but its distribution differs from that of SHG (Figs. \ref{4}(c) and \ref{5}(c)). The outer walls of the endosperms display very weak SHG signals but not 2PEF signals (data not shown).

To assign the origin of the strong SHG spot at the end of the embryo, we examined the SHG response of possible SHG active substances in the embryo. Commercially available powders of three substances (i.e., \textalpha -amylase (Kishida Chemical Co., Ltd., product code: 260-04412), glucose (Wako Pure Chemical Industries, Ltd., product code: 049-00591), and maltose (ChromaDex, Inc., product code: ASB-00013055-001) crystalline powders) as delivered were observed by our SHG microscope. All of these are SHG active (data not shown).

We also examined the SHG response of testa powder taken from the rice grains by a commercial \textquotedblleft rice huller\textquotedblright, which removes only the testa and a part of the embryo. The starch iodine reaction confirms that the peeled powder lacks starch. However, the peeled powder shows an SHG response (data not shown). We put three doses of the peeled powder of 2.5 mg each into 50 mL of water, $N,N$-dimethylformamide (DMF), and ethanol, stirred them well for 5 min, and filtered them. Three residues on the filter paper were collected and dried for 24 h.  Table \ref{t1} shows the SHG responses of the dried residues. The SHG activity of the testa powder is lost after washing with water and DMF but remains after washing with ethanol.

\begin{table*}[h]
\centering
\caption{Solubilities, Residue Components, and SHG Activities of the Residues for the Three Candidates}
\label{t2}
\begin{tabular}{llll}
\hline
 & water & DMF & ethanol \\
\hline
(i) \textalpha -amylase & soluble &insoluble& insoluble \\
(ii) glucose, maltose & soluble & soluble & insoluble \\
(iii) leukoplasts & insoluble & unknown & soluble (only lipids) \\
\hline
possible residue & (iii) & (i), (iii) & (i), (ii), (iii) [without lipids] \\
\hline
SHG of residue in Table \ref{t1} & no        & no & yes \\  
\hline
\end{tabular}
\end{table*}

\section{Discussion}
\label{sec:Discussion}
\subsection{SHG Images of EFR}
\label{sec:SHG Images of EFR}

A previous study on the cross section of a glutinous rice grain around the crush cell layer between the endosperm and the embryo reported an enhanced SFG signal at wavenumbers of 2920 and 2970 cm$^{-1}$ of the IR light \cite{li2012_3}.  This IR frequency is the resonant C-H stretching vibration region of the amylopectin molecules. A similar enhanced SHG in the glutinous rice occurs around the crush cell layer in this study (Figs. \ref{2}(e, f)). Additionally, the SHG intensity distribution profile of Fig. \ref{2}(f) is similar to the SFG profile of the previous report \cite{li2012_3}.  

There are two possibilities for the enhanced second-order nonlinear effect. One is the high density of amylopectin near the crush cell layer. The other is the highly oriented amylopectin near the crush cell layer. The iodine starch reaction (Fig. \ref{2}(c)) indicates that the density of amylopectin is rather constant in this region. Since this study reveals that both SFG and SHG are strong near the crush cell layer, we speculate that the origin of the enhanced SFG signal in the crush cell layer of the glutinous rice \cite{li2012_3} is the high degree of asymmetry in the amylopectin chain structure.

On the other hand, the nonglutinous rice does not show SHG near the crush cell layer (Figs. \ref{3}(e, f)), suggesting that crush cell layer of the nonglutinous rice has a lower concentration of amylopectin. The starch in the glutinous rice is composed mostly of amylopectin, whereas the nonglutinous rice contains amylose and amylopectin. The starch iodine reaction is rather strong near the crush cell layer of the nonglutinous rice (Fig. \ref{3}(d)).  Therefore, one possibility is that the endosperm near the crush cell layer of the nonglutinous rice has a higher concentration of amylose. Amylose has a better water absorption than insoluble amylopectin. Because the crush cell layer serves as a water reservoir in germination, the amylose content may be high in the nonglutinous rice. This hypothesis is consistent with the weak SHG in the crush cell layer of the nonglutinous rice. These results suggest that the distribution of starch in the vicinity of the crush cell layer depends on the type of rice.

In fact, we measured the vibrational spectrum of a cross section of a rice grain with the Fourier transform infrared spectroscopy (FT-IR), but the factor analysis of the infrared spectra of amylose, amylopectin, and rice starch was unsuccessful. FT-IR is a general method to distinguish between molecular species, but it is unsuited to identify starch types \cite{FTIR_starch_similar}.  The molecular units of the polysaccharides in starch and cellulose are almost the same. It has also been reported that the infrared spectroscopy of amylose, amylopectin, and rice starch have similar factor analysis patterns \cite{starch_IR}.  On the other hand, it has been reported that amylose gives a Raman band at 1657 cm$^{-1}$ while amylopectin does not \cite{Raman_1657_amylose}.  Therefore, Raman spectroscopy and microscopy can be another way to map amylose in the rice grain. The comparison of the performances of SHG and Raman spectroscopy is our future target.

2PEF images of chromophores in a young leaf tissue have been reported \cite{2PEF_plant_1,2PEF_plant_2} and the origin of the resonant 2PEF was assigned to proplastid. However, it is difficult to judge that the observed SHG in our study is due to proplastid because its resonant wavelength does not match our light source. The resonance bands of colorless saccharides in the visible region are quite unlikely, and the origin of the 2PEF in this study is not clear.

\subsection{SHG Images in Embryos}

We considered three possible origins of the SHG spots in Figs. \ref{2}(e), \ref{3}(e), \ref{4}(c) and \ref{5}(c) at the ends of the embryos: (i) amylase, (ii) crystals of glucose, maltose, or both, and (iii) leucoplasts. (i) Amylase may be stored in the aleurone layer of ungerminated rice grains \cite{okamoto1979_63}.  \textalpha -Amylase reagent has a detectable SHG signal. Because \textalpha -amylase forms a single crystal in the P2$_1$2$_1$2$_1$ space group \cite{amylase} with a broken inversion symmetry, it should have a nonzero second-order nonlinear susceptibility. (ii) Glucose and maltose are believed to be present in ungerminated grain embryos \cite{takahashi1962_14,murata1968_43}.  Detection of the SFG signal in the crystalline powder of glucose has been reported \cite{t1}.  We have confirmed the SHG response from the crystalline powder of glucose and maltose. (iii) Leucoplast is a generic term for amyloplast, elaioplast, and proteinoplast \cite{debarre2006_3}.  Since amyloplast, which can generate SHG light, is generally contained in the root cell of the plant \cite{saito2005_17}, it may also be contained in the hypocotyl and the radicle in the grain in question. Elaioplast is an organelle that stores lipids. Some lipids have optical nonlinearity \cite{debarre2006_3}. Similarly, proteinoplast is an organelle. It is only found in plant cells, especially in roots and seeds. It contains some protein crystals \cite{vigil1985_83,wise2007_2007}.  In a protein crystal, the protein molecules may show a strong second-order nonlinear response due to their well-oriented structure \cite{fine1971_10}.

The SHG observations in Table \ref{t1} are used to discuss the feasibility of the three candidates. Table \ref{t2} lists the solubility of the three candidates in the solvent. After washing with water, only (iii) leucoplast remains in the residue. The SHG signal is not observed in the water washed residue. Thus, candidate (iii) can be excluded as the origin of the strong SHG spots in embryos near embryo testa in Figs. \ref{2}(e), \ref{3}(e), \ref{4}(c) and \ref{5}(c).  After washing with DMF, (i) \textalpha -amylase remains in the residue and the DMF-washed testa does not exhibit an SHG signal. Hence, candidate (i) can also be excluded. Although the solubility of leucoplast in DMF is unknown, it is excluded above. Therefore, the remaining candidate (ii) of glucose or maltose is considered to be the origin of the strong SHG spots in the embryo.

When the rice seed absorbs water during germination, the water passes through the seed coat covering the boundary between the embryo and endosperm (Figs. \ref{4}(a) and \ref{5}(a), near the black arrows) to reach the crush cell layer. The water subsequently passes through the absorption cell layer of the scutellum and moves to embryonic tissue \cite{takahashi1962_14}.  The embryo breathes by consuming its carbohydrate but not its starch \cite{takahashi1962_14,murata1968_43}.  Since the hypocotyl needs energy for growth, glucose and maltose must move to it to be consumed. In fact, since seeds sprout when they are covered by water in the ripening season, it is highly likely that glucose and maltose will move even in ripened seeds.  

Finally, although our measurement indicates that the reagent of \textalpha -amylase is SHG active, it is assumed that the amount of \textalpha -amylase present in the ungerminated rice grains is too small for SHG detection because \textalpha -amylase in rice is synthesized only during the seed germination process \cite{alpha-amylase_biosynthesis}.

\section{Conclusion}
\label{sec:Summary}

The SHG images are obtained from ungerminated glutinous rice ({\it {Oryzaglutinosa}} cv. Shintaishomochi) and nonglutinous rice ({\it {Oryzaglutinosa}} cv. Koshihikari). The observed enhancement effect of the SHG in the crush cell layer of the glutinous rice is consistent with the enhancement effect of SFG in the previous study \cite{li2012_3}. The origin of the enhancement effect is broken centrosymmetry, not the resonance of molecular vibration. On the other hand, the enhancement effect of SHG is not observed near the crush cell layer of the nonglutinous rice. Considering the results of the starch iodine reaction, amylopectin in the crush cell layer in the glutinous rice grains may have a highly oriented structure, while the endosperm near the crush cell layer of the nonglutinous rice may have a higher concentration of amylose. Hence, the type and distribution of starch in the vicinity of the crush cell layer may depend on the type of rice. Additionally, some strong SHG spots are observed along the testa side on the embryo. However, the testa side lacks starch. Thus, the origin of these SHG spots is likely glucose, maltose, or both. In the testa side of the embryo, crystallized glucose or maltose is detected by SHG.

This study examined typical rice species. As the results demonstrate that SHG can monitor the distribution of sugars and amylopectin in the embryo and neighboring regions of rice grains, we plan to investigate more varieties to verify whether the tendencies observed here apply to rice in general.

\vspace{9mm}
\leftline{{\bf{Notes}}}
The authors declare no competing financial interest.

\bibliographystyle{0.zidingyi}
\bibliography{rice_ref}

%%%%%%%%%%%%%%%%%%%%%%%%%%%%%

\end{document}